\documentclass[aps,prd,showpacs,twocolumn,groupedaddress]{revtex4}
\usepackage{graphicx}
\usepackage{amsfonts}
\usepackage{amssymb}
\usepackage{ulem}

\begin{document}

\title{
Instantaneous interquark potential in generalized Landau gauge
in SU(3) lattice QCD: a possible gauge for the quark potential model
}
\author{Takumi~Iritani}
\affiliation{Department of Physics, Graduate School of Science, 
Kyoto University, \\
Kitashirakawa-oiwake, Sakyo, Kyoto 606-8502, Japan}
\author{Hideo~Suganuma}
\affiliation{Department of Physics, Graduate School of Science, 
Kyoto University, \\
Kitashirakawa-oiwake, Sakyo, Kyoto 606-8502, Japan}
\date{\today}

\begin{abstract}

We investigate ``instantaneous interquark potential'', 
an interesting gauge-dependent quantity defined from the spatial correlator 
$\langle \mathrm{Tr} [U_4^\dagger(s)U_4(s')]\rangle$ 
of the temporal link-variable $U_4$, in detail in generalized Landau gauge 
using SU(3) quenched lattice QCD.
While the instantaneous potential has no linear part in the Landau gauge, 
in the Coulomb gauge, it is expressed by the Coulomb plus linear potential, 
where the slope is 2-3 times larger than the physical string tension, 
and the lowest energy state is considered to be a gluon-chain state. 
Using the generalized Landau gauge, 
we find that the instantaneous potential can be continuously described 
between the Landau and the Coulomb gauges, and it approximately reproduces 
the physical interquark potential in a specific intermediate gauge,
which we call ``$\lambda_C$-gauge''.
This $\lambda_C$-gauge is expected to provide 
a quark-potential-model picture, where dynamical gluons do not appear.
We also investigate $T$-length terminated Polyakov-line correlator 
and its corresponding ``finite-time potential'' in generalized Landau gauge.
\end{abstract}



\maketitle

\section{Introduction}

Nowadays, quantum chromodynamics (QCD) is established as 
the fundamental theory of the strong interaction, 
and perturbative QCD gives a standard framework to describe 
high-energy reactions of hadrons. 
QCD is a nonabelian gauge theory constructed from quarks and gluons, 
and color SU(3) gauge symmetry is one of the guiding principles 
in formulating QCD \cite{Nambu66,GWP73}.
For actual perturbative calculations of QCD, 
the gauge has to be fixed to remove gauge degrees of freedom.
In the low-energy region, however, QCD exhibits a strong-coupling nature 
and the resulting nonperturbative QCD is a very difficult and complicated theory.

On the other hand, the quark potential model has been phenomenologically used 
for the description of low-energy properties of hadrons 
in terms of their underlying structure. 
The quark potential model is a successful nonrelativistic or semi-relativistic 
framework with a potential instantaneously acting among quarks, 
and describes many hadron properties in terms of quark degrees of freedom. 
In this model, there are no dynamical gluons, and gluonic effects 
indirectly appear as the instantaneous interquark potential.
(Here, the concept of ``instantaneous'' is to be considered 
at the scale of the effective model.)

In spite of a great success of the quark potential model, 
its relation to QCD is not yet so clear, and to link the quark potential model 
from QCD is one of the important subjects in hadron physics \cite{Isgur8586, BPSV0005,TS0304}.
In principle, the quark model may be obtained from QCD after integrating out 
the gluon degrees of freedom in the path integral formalism.
Or, from the viewpoint of ``gauge'' in QCD, 
the quark model without dynamical gluons may be regarded 
as a gauge-fixed effective theory of QCD.
In fact, the quark potential model does not have local color SU(3) symmetry 
but has global color SU(3) symmetry \cite{Isgur8586,HanNambu}, 
since each quark has a color and dynamical gluons are absent in this model framework.

If so, what gauge of QCD corresponds to the quark model?
Since the quark model has global color SU(3) symmetry and 
spatial-rotation symmetry, one may first consider that 
the Landau gauge and the Coulomb gauge would be 
candidates of such a gauge for the quark model.
Actually, the Coulomb gauge is defined by minimizing 
``spatial gluon-field fluctuations'' in total, 
as will be shown in Sec.II, and therefore one may expect 
only a small gluon-field fluctuation appearing in the Coulomb gauge.
Similarly, the Landau gauge is defined by minimizing 
``total gluon-field fluctuations'' in Euclidean QCD, 
so that only a small gluon-field fluctuation may 
be expected in the Landau gauge.
Such a minimal gluonic content in these gauges seems to 
be preferable for the modeling only with quarks.

In this paper, we investigate the instantaneous interquark potential 
in the Landau, the Coulomb, and their intermediate gauges, 
i.e., ``generalized Landau gauge'' (or ``$\lambda$-gauge''), in SU(3) lattice QCD, 
from the viewpoint of the quark potential model \cite{Iritani10}.
Here, the generalized Landau gauge is a natural general gauge to connect 
the Landau, the Coulomb, and the temporal gauges, 
by one real parameter $\lambda$.

Besides the quark-model arguments, it is meaningful to investigate 
the connection between the Landau and the Coulomb gauges, 
using the generalized Landau gauge.
Actually, these gauges have been often used as the typical gauge in QCD, 
but their physical pictures seem to be rather different 
for several important arguments in QCD.

  As the typical example, the color confinement, which is an important 
  gauge-invariant QCD phenomenon, can be explained from various viewpoints in various gauges.
  In the Landau gauge, the color confinement is mathematically investigated
  by the Kugo-Ojima criterion, in terms of the BRST charge and the inverse Higgs theorem 
  \cite{KugoOjima}.
  In the Coulomb gauge, the color confinement is argued from the viewpoint 
  of a large instantaneous Coulomb energy \cite{Gribov, Zwanziger98, Greensite03, Greensite04}, 
  and its resulting gluon-chain picture \cite{GreensiteThorn,tHooft}.
  Taking the maximally Abelian (MA) gauge, the quark confinement has been discussed 
  in terms of the dual superconductor picture \cite{NambutHooftMandelstam}.

  Of course, in gauge theories, 
  the physical quantities never depend on the gauge choice.
  However, according to the gauge choice, the physical picture can be changed,
  and the role of the gauge field, which is fundamental field of the gauge
  theory, can be also changed. Then, it is important to link the different gauges,
  and investigate the role of gluons in each gauge.
  The role and the properties of gluons are expected to be clarified by the
  overview on the structure of gauge dependence.

  One of the aim in this paper is to investigate the gluonic properties through
  a continuous view from the Landau gauge to the Coulomb gauge,
  using generalized Landau gauge. In particular, we clarify 
  the behavior of the instantaneous potential, 
  as an interesting gluonic correlation. 

  The organization of this paper is as follows.
  In Sec.II, we briefly review the properties of the Landau gauge and the Coulomb gauge.
  In Sec.III, we give the formalism of generalized Landau gauge ($\lambda$ gauge).
  In Sec.IV, we formulate the instantaneous potential in lattice QCD.
  In Sec.V, we show the lattice QCD results.
  In Sec.VI, we investigate Polyakov-line correlators and its relation to the potential.
  Sec.VII will be devoted to Summary and Discussions.

\section{Landau gauge and Coulomb gauge}
  In this section, we briefly review the properties of the Landau gauge and the Coulomb gauge.

  \subsection{Landau gauge}

  The Landau gauge is one of the most popular gauges in QCD, 
  and its gauge fixing is given by 
    \begin{equation}
      \label{eqLandauFix}
      \partial_\mu A_\mu = 0,
    \end{equation}
  where
    $A_\mu(x) \equiv A_\mu^a(x) T^a \in
    \mathfrak{su}(N_c)$ are gluon fields, 
    with $\mathfrak{su}(N_c)$ generator $T^a (a = 1,2,\dots N_c^2-1)$.
  The Landau gauge keeps the Lorentz covariance
  and the global SU($N_c$) color symmetry.
    These symmetries simplify the tensor structure of 
    various quantities in QCD.
    For example, the gluon propagator $D_{\mu\nu}^{ab}(p)$
    is simply expressed as
    \begin{equation}
      D_{\mu\nu}^{ab}(p) = D(p^2) \delta^{ab} \left( 
      g_{\mu\nu} - \frac{p_\mu p_\nu}{p^2} \right),
    \end{equation}
    due to the symmetries and the transverse property.
    Owing to this advanced feature, the Landau gauge 
   is often used both in the Schwinger-Dyson formalism
  \cite{HigashimaMiransky,Alkofer01} and in 
   lattice QCD studies for quarks and gluons \cite{Mandula99,Iritani09}.

    In Euclidean QCD, the Landau gauge has 
    a global definition to minimize the global quantity,
    \begin{equation}
      R_{\rm L} = \int d^4x \mathrm{Tr} \left\{ A_\mu(x) A_\mu(x) \right\}
        = \frac{1}{2}\int d^4x A_\mu^a(x) A_\mu^a(x),
    \end{equation}
    by the gauge transformation.
    This global definition is more strict, 
    and the local form in Eq.(\ref{eqLandauFix}) can be obtained 
    from the minimization of $R_{\rm L}$.
    Since the quantity $R_{\rm L}$ physically means the total amount of gauge-field fluctuations, 
    and therefore the Landau gauge maximally suppresses  
    artificial gauge-field fluctuations
    originated from the gauge degrees of freedom.

  Here, we comment on non-locality of the gauge fields.
  Through the gauge fixing procedure,
  gauge fields have non-locality stemming from the Faddeev-Popov determinant.
  In the Landau gauge, this non-locality of gauge fields
  is Lorentz covariant.

  Using the Landau gauge, or a covariant and globally symmetric gauge,  
  the color confinement has been mathematically investigated in terms of the BRST charge 
  and the inverse Higgs theorem, which is known as the ``Kugo-Ojima criterion'' \cite{KugoOjima}.

  \subsection{Coulomb gauge}

  The Coulomb gauge is also a popular gauge in QCD, 
  and its gauge fixing is given by
    \begin{equation}
      \partial_i A_i = 0.
    \end{equation}
  This condition resembles the Landau gauge
  condition (Eq.(\ref{eqLandauFix})),
  but there are no constraints on $A_0$.
  In the Coulomb gauge, the Lorentz covariance is partially broken, 
  and gauge field components are completely decoupled into two parts, 
  $\vec{A}$ and $A_0$:
  $\vec{A}$ behave as canonical variables and 
  $A_0$ becomes an instantaneous potential.

    Similarly in the Landau gauge, the Coulomb gauge has 
    a global definition to minimize the global quantity
    \begin{equation}
      R_{\rm Coul} \equiv \int d^4 x \mathrm{Tr} 
        \left\{ A_i(x) A_i(x)\right\}
      = \frac{1}{2} \int d^4x A_i^a(x) A_i^a(x)
    \end{equation}
    by the gauge transformation.
    Note here that 
    the Euclidean metric is not necessary for
    the global definition of the Coulomb gauge.
    Note also that there appears no nonlocality in the temporal direction 
    in the Coulomb gauge. Due to this feature, 
    a hadron mass measurement can be safely performed using 
    a spatially-extended quark source in the Coulomb gauge 
    in lattice QCD calculations \cite{CGsource}.

    In the Coulomb gauge, one of the advantages is the compatibility
    with the canonical quantization \cite{ItzyksonZuber}.
    The QCD Hamiltonian is expressed as
    \begin{eqnarray} 
      \label{eqQcdHamiltonian}
      H &=& \frac{1}{2} \int d^3x \left(\vec{E}^{a}\cdot \vec{E}^{a}
      + \vec{B}^{a} \cdot \vec{B}^{a} \right) \nonumber \\
      & &+ \frac{1}{2} \int d^3x d^3y \rho^a(x)K^{ab}(x,y)\rho^b(y), 
    \end{eqnarray}
    where $\rho^a$ is the color charge density, $\vec{E}^a$ and $\vec{B}^a$ are 
    the color electric and magnetic field, respectively.
    Here, $K^{ab}(x,y)$ is the instantaneous Coulomb propagator \cite{Greensite03}
    defined as
    \begin{equation}
     K^{ab}(x,y) = [ M^{-1} (-\nabla^2)M^{-1}]_{xy}^{ab},
    \end{equation}
   with the Faddeev-Popov operator 
   \begin{equation}
     M^{ac} = - \partial^2 \delta^{ac} - \varepsilon^{abc}A_i^b \partial_i.
   \end{equation}

    The confinement picture in the Coulomb gauge focuses on the Coulomb energy including the inverse of $M$.
    Here, the Coulomb energy is the non-local second term of the QCD Hamiltonian (\ref{eqQcdHamiltonian}), 
    and is regarded as the instantaneous potential. 
    Near the Gribov horizon, where the Faddeev-Popov operator $M$ has zero eigenvalues \cite{Gribov}, 
    the Coulomb energy at large quark distance is expected to be largely enhanced 
    and leads to a confining interquark potential, 
    which is called as the ``Gribov-Zwanziger scenario'' \cite{Gribov,Zwanziger98}.

    As Zwanziger showed, the Coulomb energy (instantaneous potential) 
    $V_{\rm Coul}(R)$ in the Coulomb gauge 
    gives an upper bound on the static interquark potential $V_{\rm phys}(R)$
    \cite{Zwanziger03}, i.e.,
    \begin{equation}
      V_{\rm phys}(R) \leq V_{\rm Coul}(R).
    \end{equation}
    This inequality indicates that 
    if the physical interquark potential is confining 
    then the Coulomb energy $V_{\rm Coul}$ is also confining.
    Actually, lattice QCD calculations \cite{Greensite03} 
    show that the Coulomb energy (the instantaneous potential) between a quark and an antiquark
    leads to a linear potential, which characterizes the confinement.
    However, the slope of the instantaneous potential is too large, 
    i.e., $2 \sim 3$ times larger than the physical string tension, 
    and this Coulomb system turns out to be an excited state.

    As for the ground-state of the quark-antiquark system, Thorn and Greensite proposed 
    the ``gluon-chain picture'' in the Coulomb gauge \cite{GreensiteThorn}.
    In fact, to screen the large Coulomb energy between the quark and the antiquark, 
    chain-like gluons are dynamically generated between them.
    This gluon-chain is expected to give the linear potential between quarks. 
    In other words, QCD string can be regarded as a ``chain'' of gluons in the Coulomb gauge.

\section{Generalized Landau gauge}
  In this section, we investigate the ``generalized Landau gauge'', 
  or ``$\lambda$ gauge'' \cite{Bernard90}, which continuously connects
  the Landau and the Coulomb gauges.

  Since the Landau gauge and the Coulomb gauge
  are useful gauges and give different interesting pictures in QCD,
  it is meaningful to show the linkage of these gauges.
  To link these gauges,
  we generalize the gauge fixing condition (\ref{eqLandauFix}) as 
  \begin{equation}
    \label{eqlambdafix}
    \partial_i A_i + \lambda \partial_4 A_4 = 0,
  \end{equation}
  by introducing one real parameter $\lambda$ \cite{Bernard90}.
  The case of $\lambda = 1$ corresponds to the Landau gauge fixing condition,
  the Coulomb gauge is achieved at $\lambda = 0$, 
  and also temporal gauge for $\lambda \rightarrow \infty$.
  Therefore, we can analyze gauge dependence of various properties
  from the Landau gauge toward the Coulomb gauge
  by varying $\lambda$-parameter from 1 to 0.
  $\lambda$-gauge keeps the global SU($N_c$) color symmetry, but 
  partially breaks the Lorentz symmetry 
  like the Coulomb gauge, except for $\lambda$=1.

  In Euclidean QCD,
  the global definition of $\lambda$-gauge is expressed by 
  the minimization of
  \begin{equation}
  \label{eqgloballambda}
    R^{\lambda} \equiv \int d^4x \left[
      \mathrm{Tr}\left\{A_i(x) A_i(x)\right\}
    + \lambda \mathrm{Tr}\left\{A_4(x) A_4(x)\right\}
    \right]
  \end{equation}
  by the gauge transformation.
  Here, the $\lambda$-parameter controls 
  the ratio of the gauge-field fluctuations of
  $\vec{A}$ and $A_4$.

  Lattice QCD is formulated on the discretized Euclidean space-time, 
  and the theory is described with the link-variable 
  $U_\mu(x) \equiv  e^{iagA_\mu(x)}\in {\rm SU}(N_c)$, 
  with the lattice spacing $a$ and the gauge coupling constant $g$,
  instead of gauge fields $A_\mu(x) \in \mathfrak{su}(N_c)$.
  $\lambda$-gauge fixing condition is expressed
  in terms of the link-variable as the maximization of a quantity 
  \begin{equation}
    \label{eqRlambda}
    R_{\rm latt}^\lambda[U] \equiv \sum_x \left\{ \sum_i \mathrm{Re}
    \mathrm{Tr} U_i(x) + \lambda \mathrm{Re} \mathrm{Tr} U_4(x) \right\}
  \end{equation}
  by the gauge transformation of the link-variables, 
  \begin{equation}
    U_\mu(x) \rightarrow \Omega(x) U_\mu(x) \Omega^\dagger(x+\hat{\mu}),
  \end{equation}
  with the gauge function $\Omega \in \mathrm{SU}(3)$.
  In the continuum limit of $a \rightarrow 0$, this
  condition results in the minimization of $R^{\lambda}$ in Eq.(\ref{eqgloballambda}), and 
  satisfies the local $\lambda$-gauge fixing condition of Eq.(\ref{eqlambdafix}).

  Note here that the gluon-field fluctuation is strongly suppressed 
  in the generalized Landau gauge with $\lambda \ne 0$, so that 
  one can use the expansion of the link-variable  
  $U_\mu(x) \equiv e^{iagA_\mu(x)} \simeq 
  1 + iagA_\mu(x) + \mathcal{O}(a^2)$ for small lattice spacing $a$, 
  and the gluon field $A_\mu(x)$ 
  can be defined by
  \begin{equation}
  \label{eqgluon}
    A_\mu(x) \equiv 
    \frac{1}{2iag}\left[U_\mu(x)- U_\mu^\dagger(x)\right]_{\rm traceless} \in \mathfrak{su}(N_c)
  \end{equation}
  without suffering from large gluon fluctuations stemming from the gauge degrees of freedom.

\section{Polyakov-line correlators and instantaneous potential}
  In this section, we formulate Polyakov-line correlators
  and instantaneous potential in lattice QCD.

  First, we consider the Euclidean continuum theory to make clear the physical
  interpretation of the terminated Polyakov-line correlator.
  Considering the source field $J_\mu(x)$ which couples
  to the gauge field $A_\mu(x)$, the generating functional is given by
  \begin{equation}
    Z[J] \equiv \langle \ e^{i\int d^4x \mathcal{L}_{\rm int}(J)} \rangle
    = \langle \exp\{ ig \int d^4x A_\mu^a(x) J^\mu_a(x)\} \rangle.
  \end{equation}
  We consider a closed-loop current such as the Wilson loop $W(R,T)$, 
  which is constructed in a gauge-invariant manner.
  From the relation $Z[J] \propto e^{-E_JT}$, 
  the ground-state energy between static quark and antiquark pair is expressed as
  \begin{equation}
    V_{\rm phys}(R) = - \lim_{T \rightarrow \infty} \frac{1}{T} \ln \langle W(R,T) \rangle.
  \end{equation}

  Next, we consider the color current $J_\mu(x)$ of 
  a quark located at $\vec{x} = \vec{a}$ and an antiquark at $\vec{x} = \vec{b}$.
  For the case that these sources are generated at $t = 0$ and annihilated at $t = T$, 
  the color current $J_\mu(x)$ is expressed as  
  \begin{equation}
    J_\mu(x) = \delta_{\mu 4}\left[ \delta(\vec{x}-\vec{a}) - \delta(\vec{x}-\vec{b})\right]
    \theta(T-t)\theta(t),
  \end{equation}
  in the generalized Landau gauge.
  In this case, the current $J_\mu$ is not conserved, and it breaks the gauge invariance.   
  In a similar manner to the Wilson loop $W(R,T)$,
  we define the ``energy'' of two sources
  in the presence of $J_\mu(x)$ as
  \begin{equation}
    \label{eqterminatedV}
    V(R,T) = - \frac{1}{T} \ln \langle \mathrm{Tr} [L(\vec{a},T) L^\dagger(\vec{b},T)] \rangle
  \end{equation}
  with $R = |\vec{a}-\vec{b}|$,
  using the terminated Polyakov-line
  \begin{equation}
    \label{eqterminatedPL}
    L(\vec{x},T) \equiv P \exp\left(ig\int_0^T dx_4 A_4(\vec{x},x_4)\right),
  \end{equation}
  with $P$ being the path-ordered product.
  As a caution, for $\lambda \ne$ 0, $V(R,T)$ is gauge-dependent and
  does not mean the energy of a physical state, due to the temporal nonlocality of
  the Faddeev-Popov determinant.

  Next,
  we consider $N_x \times N_y \times N_z \times N_t$ lattice
  with the lattice spacing $a$.
  Using temporal link-variables, 
  the terminated Polyakov-line with length $T$ is defined as
  \begin{equation}
    L(\vec{x},T) = U_4(\vec{x},a) U_4(\vec{x},2a)\cdots U_4(\vec{x},T).
  \end{equation}
  The terminated Polyakov-line is generally gauge variant, 
  and its expectation value depends on the choice of gauge.
  Only for $T = N_t$, the trace of the Polyakov-line coincides the Polyakov loop, 
  which is gauge invariant. 
  In the Coulomb gauge, the expectation value of the terminated Polyakov-line 
  is zero due to the remnant symmetry. (See Appendix A.)

  In this paper, we consider the Polyakov-line correlator 
  in generalized Landau gauge denoted by 
  \begin{equation}
    G_\lambda(R,T)
      = \langle \mathrm{Tr} [L^\dagger(\vec{x},T)L(\vec{y},T)]\rangle
  \end{equation}
  with $R = |\vec{x} -\vec{y}|$.
  From this correlator, we define ``finite-time potential'', 
  \begin{equation}
    V_\lambda(R,T) \equiv - \frac{1}{T} \ln G_\lambda(R,T).
  \end{equation}
  Here, for the simple expression, 
  a normalization factor 1/3 is dropped off for $G_\lambda(R,T)$, 
  since it only gives a constant term in the potential $V_\lambda(R,T)$.

  Particularly for $T = a$, we call 
  \begin{equation}
    \label{eqInstantaneousPotential}
    V_\lambda(R) \equiv V_\lambda(R,a) 
    = - \frac{1}{a} \ln \langle \mathrm{Tr}
    \big[ U_4^\dagger(\vec{x},a) U_4(\vec{y},a)\big]\rangle
  \end{equation}
  as ``instantaneous potential''.

  Here, these quantities depend on $\lambda$-parameter.
  In the Coulomb gauge, the instantaneous potential $V_{\lambda=0}(R)$ 
  (or the Coulomb energy) gives a linear potential, but 
  its slope is about $2 \sim 3$ times larger than
  the physical string tension \cite{Greensite03}.
  In the Landau gauge, 
  the instantaneous potential $V_{\lambda=1}(R)$ has no linear part \cite{Nakamura06},
  which is also expected from the exponential reduction of the gluon propagator \cite{Mandula99,Iritani09}
  and the Lorentz symmetry.
  In Sec.VII, we will discuss the relation between the gluon propagator
  and the instantaneous potential in the Landau gauge.

\section{Lattice QCD results}

  We perform SU(3) lattice QCD Monte Carlo calculations
  at the quenched level.
  We use the standard plaquette action with the lattice parameter 
  $\beta \equiv \frac{2N_c}{g^2}=5.8$ on a $16^4$-size lattice.
  The lattice spacing $a$ is 0.152 fm,
  which is determined so as to reproduce the string tension as
  $\sqrt{\sigma} = 427$ MeV \cite{STI}.

  We use the gauge configurations, which are picked up every
  1000 sweeps after a thermalization of 20000 sweeps.
  After the generation of gauge configurations, 
  we perform gauge fixing by maximizing $R^{\lambda}_{\rm latt}[U]$.
  In this paper, we use the Landau gauge ($\lambda$=1),
  the Coulomb gauge ($\lambda$=0), and 
  their intermediate gauges with $\lambda = 0.75, 0.50, 0.25, 0.10,
  0.05, 0.04, 0.03, 0.02, 0.01$.
  We investigate in detail the region near the Coulomb gauge ($\lambda$=0), 
  since the behavior of the instantaneous potential 
  largely changes for $\lambda \sim $0, as will be shown later.
  The number of gauge configurations is 50 for each $\lambda$.
  We adopt the jackknife method to estimate the statistical error.

  Here, we comment on $\lambda$-gauge fixing convergence.
  We fix the gauge by maximizing the quantity $R^\lambda_{\rm latt}[U]$ 
  in Eq.(\ref{eqRlambda}), which corresponds to 
  $\partial_i A_i + \lambda \partial_4 A_4 = 0$.
  Therefore, to check the convergence of gauge fixing, we evaluate $\epsilon_\lambda$ 
  defined by
  \begin{eqnarray} 
    \epsilon_\lambda &\equiv& 
    \langle \left(\partial_i A_i^a + \lambda \partial_4 A_4^a \right)^2 \rangle \nonumber \\
    &\equiv &
    \frac{1}{(N_c^2-1){N_{\rm site}}}\sum_{x=1}^{N_{\rm site}} \sum_{a=1}^{N_c^2-1}
    \Big\{ \sum_{i=1}^3 \left[ A_i^a(x) - A_i^a(x-\hat{i})\right] \nonumber \\
    & + & \lambda \left[ A_4^a(x) - A_4^a(x-\hat{4})\right] \Big\}^2,
    \label{eqConvergence}
  \end{eqnarray}
  with the gluon field $A_\mu(x)=A_\mu^a(x)T^a$ given in Eq.(\ref{eqgluon}).
  We iterate the gauge transformation 
  to satisfy $\epsilon_\lambda < {10}^{-12}$ finally.
  As for the instantaneous potential $V_\lambda(R)$,
  this convergence condition is very strict.
  Actually, we can obtain stable the lattice data of $V_\lambda(R)$
  even with $\epsilon_\lambda < {10}^{-4}$.

  Also, we comment that the calculation cost of the gauge fixing is rapidly increasing
  as $\lambda$ approaches to zero, while the Coulomb gauge ($\lambda=0$) itself can be easily achieved.
  Considering this critical slowing down of the gauge fixing
  \cite{Bernard90,Cucchieri07},
  we adopt a relatively small-size lattice of $16^4$ with $\beta = 5.8$, 
  although its physical volume of about $(2.4 {\rm fm})^4$ 
  is large enough to extract the relevant region for the interquark potential.

  \subsection{``Instantaneous inter-quark potential'' in generalized Landau gauge}

\begin{figure}
    \centering
    \includegraphics[width=8cm,clip]
      {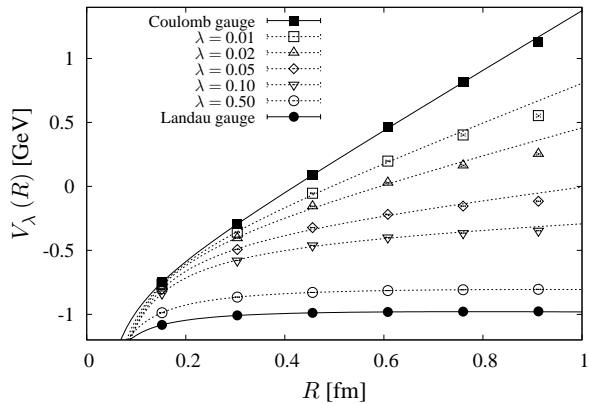}
    \caption{\label{figInstaPot}
    ``Instantaneous potential''
    $V_\lambda(R)$
    in generalized Landau gauge
    for typical values of $\lambda$.
    The symbols denote lattice QCD results, and
    the curves fit-results using Coulomb plus linear form, 
    $V_\lambda(R) = -A_\lambda/R+\sigma_\lambda R +C_\lambda$,
    in the region of $R \lesssim 0.8$ fm.
    For $\lambda = 1 \sim 0.1$, 
    the potential has almost no linear part.
    For $\lambda \lesssim 0.1$,
    the linear potential grows rapidly, and 
    $\sigma_\lambda \simeq 2.6\sigma_{\rm phys}$ at $\lambda = 0$.
    }
  \end{figure}

  We investigate the instantaneous potential 
  $V_\lambda(R)$ defined by Eq.(\ref{eqInstantaneousPotential}) 
  in generalized Landau gauge.
  Figure \ref{figInstaPot} shows lattice QCD results
  of $V_\lambda(R)$
  for typical values of $\lambda$.
  In this figure, the statistic error is small and the error bars are hidden in the symbols.

  In the Coulomb gauge ($\lambda = 0$), the instantaneous potential 
  shows linear behavior, while there is no linear part at all 
  in the Landau gauge ($\lambda = 1$).
  Thus, there is a large gap between these gauges 
  in terms of the instantaneous potential.
  In our framework, however, these two gauges are connected continuously.

  By varying the $\lambda$-parameter from 1 to 0 in the generalized Landau gauge, 
  we find that the instantaneous potential $V_\lambda(R)$ changes continuously, and  
  the infrared slope of the potential $V_\lambda(R)$ at $R \simeq 0.8{\rm fm}$ grows monotonically,
  from the Landau gauge to the Coulomb gauge, as shown in Fig.~\ref{figInstaPot}.
  Note here that the growing of the infrared slope of $V_\lambda(R)$ is quite rapid
  for $\lambda \lesssim 0.1$, near the Coulomb gauge,
  while the infrared slope is rather small and almost unchanged for $\lambda = 0.1 \sim 1$.

  To analyze the instantaneous potential $V_\lambda(R)$ quantitatively,
  we fit the lattice QCD results using 
  the Coulomb plus linear Ansatz as
  \begin{equation}
    V_\lambda(R) = - \frac{A_\lambda}{R} + \sigma_\lambda R + C_\lambda,
    \label{eqCoulplusLin}
  \end{equation}
  where $A_\lambda$ is the Coulomb coefficient and $C_\lambda$  a constant.
  Here, $\sigma_\lambda$ is the infrared slope of the potential, 
  which we call as ``instantaneous string tension''.
  Besides the Coulomb plus linear Ansatz, 
  we try several candidates of the functional form,
  $-A/R + \sigma(1-e^{-\varepsilon R})/\varepsilon$, $-A\exp(-mR)/R$,
  $-A/R + \sigma R^d$, and $-A/R^d$ apart from an irrelevant constant, 
   but they are less workable.
  The curves in Fig. \ref{figInstaPot} are the best-fit results using
  Eq.(\ref{eqCoulplusLin}). 
  The Coulomb plus linear Ansatz works well 
  at least for $R \lesssim 0.8$fm, 
  which is relevant region for hadron physics.
  We note that the Yukawa form of $-Ae^{-mR}/R$ 
  also works well near the Landau gauge, which will be discussed   
  in relation to the gluon-propagator behavior in Sec.VI-C.

  We summarize the best-fit parameters and the fit-range in Table \ref{tabFitResult}.
  While the Coulomb coefficient $A_\lambda$ has a relatively weak $\lambda$-dependence, 
  the instantaneous string tension $\sigma_\lambda$ shows a strong $\lambda$-dependence near 
  the Coulomb gauge, i.e., $\lambda \lesssim 0.1$.

  We comment on the asymptotic value of the instantaneous potential $\sigma_\lambda$.
  In the deep IR limit, $R \rightarrow \infty$, $V_\lambda(R)$ 
  goes to a saturated value, except for $\lambda = 0$.
  This feature is closely related to the property of the temporal 
  link-variable correlator $\langle \mathrm{Tr} \{ U_4^\dagger(x) U_4(y) \} \rangle$,
  as will be discussed in Sec.VI.

  \begin{table}
    \begin{center}
      \caption{\label{tabFitResult}
      Best-fit parameters on the instantaneous potential using
      $V_\lambda(R) = - A_\lambda/R + \sigma_\lambda R + C_\lambda$,
      and the ratio of the slope $\sigma_\lambda$ to the physical string tension 
      $\sigma_{\rm phys}$.
      The standard parameters of the physical interquark potential 
      are $A_{\rm phys} \simeq 0.27$ and $\sigma_{\rm phys} \simeq 0.89$GeV/fm \cite{STI}.
      The string tension $\sigma_\lambda$ is rather small for $\lambda = 0.1 \sim 1$.}
      \begin{tabular}{cccccc}
      \hline
      \hline 
      $\lambda$ & fit-range [fm] & $A_\lambda$ & $\sigma_\lambda$ [GeV/fm] 
        & $C_\lambda$ [GeV] & $\sigma_\lambda/\sigma_{\rm phys}$ \\
      \hline
      0.00 & 0.1-1.0 & 0.167(11) & 2.283(35) & -0.881(20) & 2.57(4) \\
      0.01 & 0.1-0.8 & 0.287(27) & 1.476(78) & -0.617(46) & 1.66(9) \\
      0.02 & 0.1-0.8 & 0.346(32) & 1.005(90) & -0.481(54) & 1.13(10) \\
      0.03 & 0.1-0.8 & 0.372(32) & 0.728(86) & -0.416(53) & 0.82(10) \\
      0.04 & 0.1-0.8 & 0.382(30) & 0.557(79) & -0.386(50) & 0.63(9) \\
      0.05 & 0.1-0.8 & 0.386(29) & 0.441(73) & -0.372(47) & 0.50(8) \\
      0.10 & 0.1-0.8 & 0.365(20) & 0.169(46) & -0.390(31) & 0.19(5) \\
      0.25 & 0.1-0.8 & 0.281(6) & -0.005(13) & -0.544(9) & -0.01(1) \\
      0.50 & 0.1-0.8 & 0.198(0) & -0.042(1) & -0.724(1) & -0.05(0) \\
      0.75 & 0.1-0.8 & 0.152(1) & -0.043(3) & -0.839(2) & -0.05(0) \\
      1.00 & 0.1-0.8 & 0.123(2) & -0.040(3) & -0.917(3) & -0.04(0) \\
      \hline 
      \hline
      \end{tabular}
    \end{center}
  \end{table}

\begin{figure}
    \centering
    \includegraphics[width=8cm,clip]
      {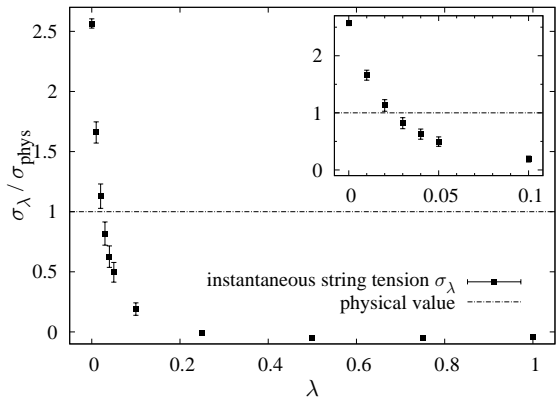}
    \caption{ \label{figStringTension}
      Instantaneous string tension $\sigma_\lambda$, the slope of 
      the linear part in the instantaneous potential $V_\lambda(R)$ 
      in the generalized Landau gauge.
      The right upper figure is a close-up near the Coulomb gauge ($\lambda$=0).
      For $0 \le \lambda \lesssim 0.1$,  
      $\sigma_\lambda$ changes rapidly from $2.6 \sigma_{\rm phys}$ to 0,
      while $\sigma_\lambda$ is rather small for $0.1 \lesssim \lambda \le 1$.
      The instantaneous string tension $\sigma_\lambda$ coincides 
      with the physical value around $\lambda_C \simeq 0.02$.
    }
  \end{figure}

  Now, we focus on the $\lambda$-dependence of 
  instantaneous string tension $\sigma_\lambda$ in Fig.~\ref{figStringTension}.
  For $0.1 \lesssim \lambda \le 1$, including the Landau gauge ($\lambda$=1), 
  $\sigma_\lambda$ is almost zero,
  so that this region can be regarded as ``Landau-like.''
  For $\lambda \lesssim 0.1$, 
  $V_\lambda(R)$ is drastically changed near the Coulomb gauge, 
  and $\sigma_\lambda$ grows rapidly in this small region.
  Finally, in the Coulomb gauge ($\lambda$=0), one finds 
  $\sigma_\lambda \simeq 2.6\sigma_{\rm phys}$, 
  with $\sigma_{\rm phys} \simeq 0.89$GeV/fm.

  Note that the instantaneous string tension 
  $\sigma_\lambda$ continuously changes from $0$ to  
  $(2 \sim 3)\sigma_{\rm phys}$, 
  according to the change from the Landau gauge to the Coulomb gauge, 
  and therefore
  there exists some specific $\lambda$-parameter of $\lambda_C \in [0,1]$
  where the slope of the instantaneous potential $V_\lambda(R)$
  coincides with the physical string tension $\sigma_{\rm phys}$. 
  Since the instantaneous potential generally depends 
  on the lattice parameter $\beta$, i.e., the lattice spacing $a$ 
  \cite{Greensite03,Iritani10}, the value of $\lambda_C$ is 
  $\beta$-dependence, although its dependence would be rather weak, 
  as will be discussed in Sec.VI.
  However, from the continuity between 
  the overconfining potential in the Coulomb gauge
  and the saturated potential in the Landau gauge,
  there must exist $\lambda_C \in [0,1]$ 
  where the instantaneous string tension $\sigma_\lambda$
  coincides with $\sigma_{\rm phys}$ for each lattice spacing.
  We call this specific gauge as ``$\lambda_C$-gauge''.
  From Fig.\ref{figStringTension}, 
  the value of $\lambda_C$ is estimated to be about 0.02 at $\beta$=5.8.
  Note here that $\lambda_C \simeq 0.02 \ll 1$ is very small,   
  and then the $\lambda_C$-gauge is close to the Coulomb gauge, 
  which indicates the small temporal non-locality. 

\begin{figure}
    \centering
    \includegraphics[width=8cm,clip]
      {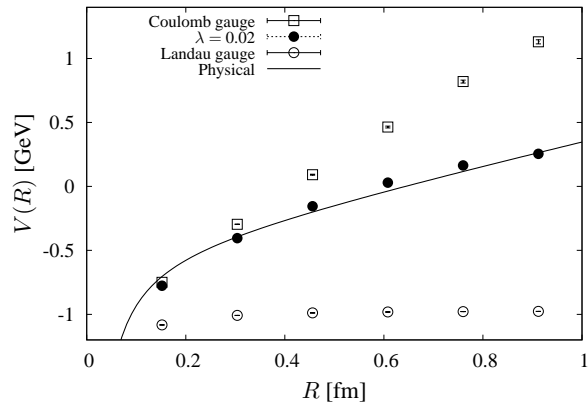}
    \caption{ \label{figPotWithCornell}
      The comparison between 
      the instantaneous potential $V_{\lambda}(R)$ at $\lambda = 0.02 (\simeq \lambda_C)$ (black dots),
      and the physical interquark potential $V_{\rm phys}(R)$ (solid line).
      In $\lambda_C$-gauge, 
      $V_{\rm phys}(R) = - \lim_{T\rightarrow \infty}\frac{1}{T}\ln
      \langle W(R,T) \rangle \simeq -A_{\rm phys}/R + \sigma_{\rm phys} R$  
      ($A_{\rm phys} \simeq$ 0.27, $\sigma_{\rm phys} \simeq$ 0.89GeV/fm)
      is approximately reproduced 
      by the instantaneous potential 
      $V_{\lambda_C}(R) = - \frac{1}{a}\ln \langle \mathrm{Tr} 
      U_4^\dagger(\vec{x},a)U_4(\vec{y},a)\rangle_{\lambda_C}$.
    }
  \end{figure}

\begin{figure}
    \centering
    \includegraphics[width=8cm,clip]{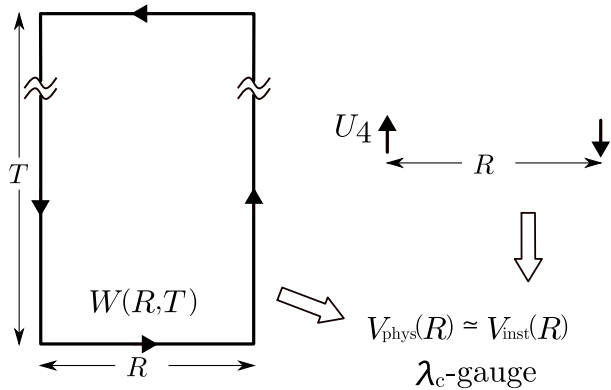}
    \caption{ 
      \label{figInstPotAndPhysPot}
      The schematic illustration of the physical interquark potential $V_{\rm phys}(R)$
      and the instantaneous potential $V_{\rm inst}(R)$.
      In $\lambda_C$-gauge, the instantaneous potential $V_{\lambda_C}(R)$
      approximately reproduces the physical potential $V_{\rm phys}(R)$.
    }
\end{figure}

  Figure \ref{figPotWithCornell} shows 
  the instantaneous potential $V_{\lambda}(R)$ at $\lambda = 0.02 \simeq \lambda_C$
  and the physical static interquark potential $V_{\rm phys}(R)$.
  In this $\lambda_C$-gauge, 
  $V_{\rm phys}(R)$ is found to be approximately reproduced  
  by $V_{\lambda_C}(R)$ for $R \lesssim$ 0.8fm. 
  While the physical static potential $V_{\rm phys}(R)$ is derived from the {\it large-T behavior} 
  of the Wilson loop $W(R,T)$ \cite{Rothe} as 
  \begin{equation}
    V_{\rm phys}(R) = - \lim_{T \rightarrow \infty}
    \frac{1}{T} \ln \langle W(R,T)\rangle,
  \end{equation}
  only {\it instantaneous corerlation} of the temporal link-variable $U_4$
  approximately reproduces $V_{\rm phys}(R)$ in the $\lambda_C$-gauge, i.e., 
  \begin{eqnarray}
    V_{\rm phys}(R) 
    &\simeq& V_{\lambda_C}(R)  \nonumber \\
    &=& - \frac{1}{a} \ln \langle 
    \mathrm{Tr} U_4^\dagger(\vec{x},a) U_4(\vec{y},a) \rangle_{\lambda_C},
  \end{eqnarray}
  as is schematically illustrated in Fig.\ref{figInstPotAndPhysPot}.

 On the relation to the confinement, which is a gauge independent phenomenon, 
 the role of gluons generally depends on the choice of gauges,
 and the physical picture of the confinement would be changed according to gauges.
 For example, in the Coulomb gauge, the instantaneous Coulomb energy gives an overconfining potential, 
 and the ground-state of the quark-antiquark system is described as the gluon-chain state. 
 On the other hand, in the Landau gauge, the instantaneous potential has no linear part, and 
 the ghost behavior in the deep-infrared region would be more important for the confinement.

  In the $\lambda_C$-gauge, the physical interquark potential
  $V_{\rm phys}(R)$ is approximately reproduced 
  by the instantaneous potential $V_{\lambda_C}(R)$.
  This physically means that all other complicated effects 
  including dynamical gluons and ghosts are approximately cancelled
  in the $\lambda_C$-gauge, and therefore we do not need to 
  introduce any redundant gluonic degrees of freedom. 
  The absence of dynamical gluon degrees of freedom 
  would be a desired property for the quark model picture. 

  \subsection{``Finite-time potential'' and ``finite-time string tension''}

    In the previous subsection, we investigated 
    the instantaneous potential $V_\lambda(R)$, which is defined
    by the Polyakov-line correlator with a minimum length on the lattice.
    For the quark-potential model, it is desired that 
    the interquark potential does not have large dependence 
    on the temporal length $T$ of the typical reaction scale.
    From this viewpoint, 
    we investigate the ``finite-time potential'' $V_\lambda(R,T)$ defined by Eq.(\ref{eqterminatedV}) in Sec.IV,
    and its temporal-length dependence.  
    Here, $V_\lambda(R,T)$ is expressed by $T$-length terminated Polyakov-line 
    $L(\vec{x},T)$ in Eq.(\ref{eqterminatedPL}),
    and a generalization of the instantaneous potential $V_\lambda(R)$.

\begin{figure}
      \centering
      \includegraphics[width=8cm,clip]
      {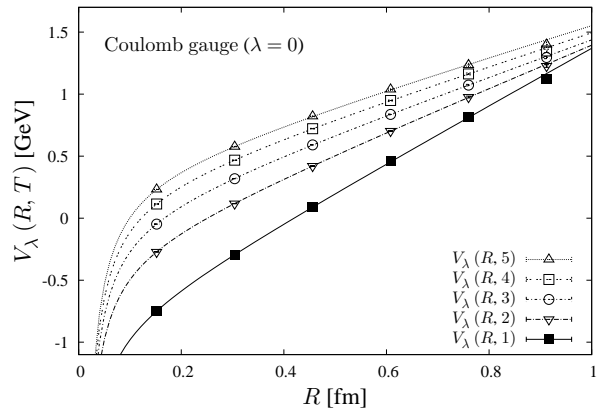}
      \caption{\label{figExtendedPot}
        ``Finite-time potential'' $V_\lambda(R,T)$
        in the Coulomb gauge ($\lambda = 0$) for $T$=1,2,3,4,5.
        Here, for the comparison, an irrelevant constant is shifted for each $T$.
        The curves denote the fit-results using the Coulomb plus linear form.
        The slope of $V_\lambda(R,T)$ is clearly changed according to $T$.
      }
    \end{figure}

  First, we consider the Coulomb gauge \cite{Greensite03,Iritani10}.
  Figure \ref{figExtendedPot} shows 
  the lattice QCD result for $V_\lambda(R,T)$ in the Coulomb gauge. 
  Similar to the instantaneous potential,
  $V_\lambda(R,T)$ is well reproduced by the Coulomb plus linear form.
  However, the parameter values are changed according to $T$-length.
  In particular, the slope of the potential becomes smaller 
  as $T$ becomes larger, which shows an ``instability'' of 
  $V_\lambda(R,T)$ in terms of $T$ in the Coulomb gauge.

  For general $\lambda$, 
  finite-time potential $V_\lambda(R,T)$ is found to be reproduced 
  by the Coulomb plus linear form as 
  \begin{equation}
    V_\lambda(R,T) 
    = - \frac{A_\lambda(T)}{R} + \sigma_\lambda(T) R + C_\lambda(T),
  \end{equation}
  at least for $R \lesssim 0.8$fm, similarly for the instantaneous potential. 

\begin{figure}
      \centering
      \includegraphics[width=8cm,clip]
      {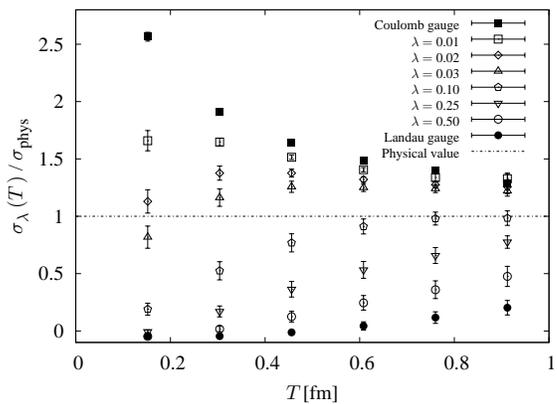}
      \caption{\label{figStringTensionLengthDep}
      $T$-length dependence of ``Finite-time string tension'' $\sigma_\lambda(T)$, 
      the infrared slope of finite-time potential $V_\lambda(R,T)$, 
      in generalized Landau gauge for several typical $\lambda$-values.
      Near the Coulomb gauge, e.g., for $\lambda \lesssim 0.03$,
      $\sigma_\lambda(T)$ goes to the same value for large $T \sim$ 1fm.
      For $\lambda \gtrsim 0.1$, $\sigma_\lambda(T)$ is an increasing function of $T$.
      In fact, even though the instantaneous potential has no linear part, 
      the linear part of $V_\lambda(R,T)$ appears and gradually grows, as the Polyakov-line grows.
      }
    \end{figure}

  Since our main interest is linear part of the potential, 
  we focus on ``finite-time string tension'' $\sigma_\lambda(T)$, 
  the slope of $V_\lambda(R,T)$.
  Figure \ref{figStringTensionLengthDep} shows 
  $\sigma_\lambda(T)$ in generalized Landau gauge
  for typical values of $\lambda$.
    In Table \ref{tabTlengthStringTension}, 
    we summarize the best-fit parameters of 
    $\sigma_\lambda(T)$ at $T = 1, 2, \dots, 6$,
    and the ratio of 
    $\sigma_\lambda(1)/\sigma_\lambda(6)$,    
    $\sigma_\lambda(1)/\sigma_{\rm phys}$, and
    $\sigma_\lambda(6)/\sigma_{\rm phys}$, respectively.

    \begin{table*}
      \begin{center}
        \caption{\label{tabTlengthStringTension}
        Finite-time string tension $\sigma_\lambda(T)$ in generalized Landau gauge
        for $T = 1, 2, \dots, 6$, together with the ratio,  
        $\sigma_\lambda(1)/\sigma_\lambda(6)$,    
        $\sigma_\lambda(1)/\sigma_{\rm phys}$, and
        $\sigma_\lambda(6)/\sigma_{\rm phys}$.
        The fit-range is the same as that listed in Table \ref{tabFitResult}.
        }
        \begin{tabular}{cccccccccc}
          \hline
          \hline
                    & \multicolumn{6}{c}{$\sigma_\lambda(T)$ [GeV/fm]} \\
          $\lambda$ & $T=1$ & $T=2$ & $T=3$ & $T=4$ & $T=5$ & $T=6$ &
              $\sigma_\lambda(1)/\sigma_\lambda(6)$ &
              $\sigma_\lambda(1)/\sigma_{\rm phys}$ &
              $\sigma_\lambda(6)/\sigma_{\rm phys}$ \\
          \hline
0.00  &  2.283(35) & 1.704(11) & 1.463(8) & 1.322(16) &  1.244(24) & 1.147(75) & 1.99(13) &   2.57(4) &  1.29(8) \\
0.01  &  1.476(78) & 1.466(30) & 1.348(17) & 1.252(16) &  1.191(27) & 1.181(43) & 1.25(8) &   1.66(9) &  1.33(5) \\
0.02  &  1.005(90) & 1.225(55) & 1.225(31) & 1.176(20) &  1.135(25) & 1.119(53) & 0.90(9) &   1.13(10)&  1.26(6) \\
0.03  &  0.728(86) & 1.034(67) & 1.119(43) & 1.113(30) &  1.104(30) & 1.086(40) & 0.67(8) &   0.82(10)&  1.22(5) \\
0.04  &  0.557(79) & 0.896(72) & 1.030(51) & 1.057(35) &  1.053(30) & 1.019(53) & 0.55(8) &   0.63(9) &  1.15(6) \\
0.05  &  0.441(73) & 0.785(76) & 0.947(59) & 1.002(43) &  1.021(35) & 1.020(44) & 0.43(7) &   0.50(8) &  1.15(5) \\
0.10  &  0.169(46) & 0.467(71) & 0.684(70) & 0.811(58) &  0.872(50) & 0.875(56) & 0.19(5) &   0.19(5) &  0.98(6) \\
0.25  &  -0.005(13) & 0.152(42) & 0.324(60) & 0.474(65) &  0.586(62) & 0.690(48) & -0.01(2) & -0.01(1) &  0.78(5) \\
0.50  &  -0.042(1) & 0.015(21) & 0.111(41) & 0.218(57) &  0.320(69) & 0.423(79) & -0.10(2) & -0.05(0)  &  0.48(9) \\
0.75  &  -0.043(3) & -0.025(11) & 0.028(27) & 0.100(44) &  0.181(58) & 0.275(69) & -0.16(4) & -0.05(0)  &  0.31(8) \\
1.00  &  -0.040(3) & -0.041(5) & -0.012(18) & 0.039(32) &  0.104(45) & 0.180(56) & -0.22(7) & -0.04(0) &  0.20(6) \\
          \hline
          \hline
        \end{tabular}
      \end{center}
    \end{table*}

    On the $T$-dependence of the finite-time string tension $\sigma_\lambda(T)$,
    there are three groups of the $\lambda$-parameter region:
    (i) Coulomb-like region ($0 \le \lambda \ll \lambda_C$),
    (ii) Landau-like region ($\lambda_C \ll \lambda \le 1$), and  
    (iii) $\lambda_C$-like region ($\lambda \sim \lambda_C$).

(i) The first category is the Coulomb-like region of $0 \le \lambda \ll \lambda_C$, 
    i.e., $0 \le \lambda \lesssim 0.01$. 
    In this region, the instantaneous string tension $\sigma_\lambda \equiv \sigma_\lambda(T=1)$
    is larger than the physical string tension $\sigma_{\rm phys}$, and 
    the instantaneous potential $V_\lambda(R) \equiv V_\lambda(R,T=1)$ gives an overconfining potential. 
    The ground-state of the quark-antiquark system is considered as the gluon-chain state.
    As the temporal length $T$ of the Polyakov-line increases, finite-time string tension 
    $\sigma_\lambda(T)$ decreases and approaches to the physical string tension $\sigma_{\rm phys}$.
    This decreasing behavior is interpreted that 
    the component of the ground-state, i.e., the gluon-chain state, 
    becomes dominant as $T$-length becomes large.
    Thus, this region would not be compatible with the quark potential picture, 
    because of the large $T$-dependence of the confining force $\sigma_\lambda(T)$ 
    in addition to the dynamical generation of the gluon-chain.

(ii) The second category is the Landau-like region of $\lambda_C \ll \lambda \le 1$, 
     i.e. $0.1 \lesssim \lambda \le 1$. 
     In this region, the instantaneous string tension is almost zero, 
     i.e., $\sigma_\lambda \simeq 0$,
     and finite-time string tension $\sigma_\lambda(T)$ is an increasing function of $T$.
     Although its asymptotic value is unclear for $T \sim 0.8$ fm,
     $\sigma_\lambda(T)$ seems to approach to the physical string tension $\sigma_{\rm phys}$,
     which will be discussed in Sec.VI.

\begin{figure}
      \centering
      \includegraphics[width=8cm,clip]
      {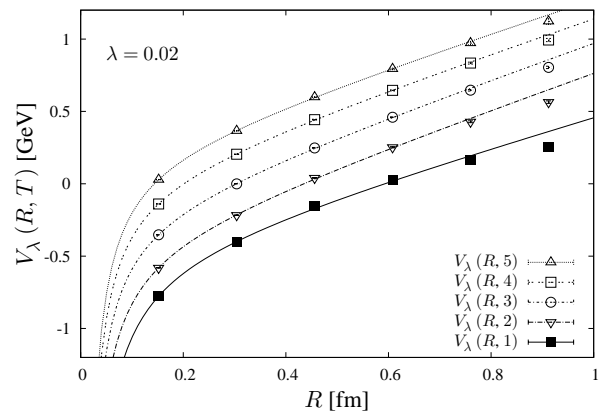}
      \caption{\label{figExtendedPotLambdaC}
      Finite-time potential $V_\lambda(R,T)$ at $\lambda = 0.02(\simeq \lambda_C)$.
      For the comparison, an irrelevant constant is shifted for each $T$.
      The slope of $V_\lambda(R,T)$ is almost the same for $T=1,2, \cdots 5$, and thus 
      the shape of $V_\lambda(R,T)$ is rather stable against the temporal length $T$.
      }
    \end{figure}

(iii) The third category is the $\lambda_C$-like region of $\lambda \sim \lambda_C$, 
      i.e., $0.01 \lesssim \lambda \lesssim 0.1$.
      In this region, the instantaneous string tension $\sigma_\lambda$ 
      is approximately equal to the physical string tension, i.e., 
      $\sigma_\lambda \simeq \sigma_{\rm phys} (\simeq$ 0.89GeV/fm), 
      and the instantaneous potential $V_\lambda(R)$ approximately reproduces 
      the physical static potential $V_{\rm phys}(R)$, 
      as shown in Fig.~\ref{figPotWithCornell}.
      As the temporal length $T$ increases, 
      finite-time string tension $\sigma_\lambda(T)$  
      is slightly changed and takes a little larger value ($\simeq$ 1.1GeV/fm) 
      around $T \simeq$ 0.8fm.
      In particular, near $\lambda_C \simeq 0.02$, 
      $\sigma_\lambda(T)$ shows only a weak $T$-dependence,   
      while $\sigma_\lambda(T)$ largely changes as $T$ in the Coulomb gauge. 
      As a whole, finite-time potential $V_\lambda(R,T)$ has small $T$-dependence, 
      as shown in Fig.~\ref{figExtendedPotLambdaC}.
      This is also a desired feature for the linkage to the quark potential model.

    When the length $T$ of the Polyakov-line increases,
    $\lambda$-dependence of $\sigma_\lambda(T)$ is weakened,
    and $\sigma_\lambda(T)$ seems to converge to  
    the physical string tension $\sigma_{\rm phys}$ for enough large $T$,
    as indicated in Fig.\ref{figStringTensionLengthDep}.

    There are two ingredients on the above gauge-dependence ($\lambda$-dependence):
    one is the large excess of the Coulomb energy in the Coulomb gauge,   
    the other is the non-locality from the Faddeev-Popov determinant.
    From the fixing condition of generalized Landau gauge 
    in Eq.(\ref{eqlambdafix}), one finds that 
    the $\lambda$-parameter controls the non-locality in the temporal direction.
    In the Landau gauge, the non-locality appears 
    equally in spatial and temporal directions,
    while temporal non-locality disappears in the Coulomb gauge.
    If the Coulomb-energy excess can be neglected, e.g., for large $\lambda$, 
    $V_\lambda(R,T)$ is expected to reproduce the static potential $V_{\rm phys}(R)$,
    when $T$-length and $R$ are large enough to neglect the non-locality scale. 
    Owing to the $\lambda$-dependence of the non-locality, 
    such a $T$-length exceeding the non-locality is to be larger 
    for larger $\lambda$ in the Landau-like region.

    Near the $\lambda_C$-gauge, 
    finite-time potential $V_\lambda(R,T)$ has only weak $T$-length dependence.
    In other words, this can be regarded as an approximate ``fixed point'' against $T$
    around $\lambda_C \simeq 0.02$.
    We conjecture that this is due to the approximate cancellation between 
    the Coulomb-energy excess and the non-locality.
    Actually, in contrast to the large $T$-dependence of $V_\lambda(R,T)$ 
    in the Coulomb gauge as shown in Fig.\ref{figExtendedPot}, 
    $V_{\lambda_C}(R,T)$ is rather stable against $T$-length in the $\lambda_C$-gauge, 
    as shown in Fig.\ref{figExtendedPotLambdaC}.

\section{Terminated Polyakov-line correlator and potentials}

  In the previous section, 
  we investigated instantaneous potential $V_\lambda(R)$
  and finite-time potential $V_\lambda(R,T)$, which are derived from 
  the correlation of terminated Polyakov-line $L(\vec{x},T)$.
  In this section, we investigate properties of the Polyakov-line correlator, 
  and clarify its relation to $V_\lambda(R)$ and $V_\lambda(R,T)$.

  \subsection{Asymptotic behavior of link-variable correlator and instantaneous potential}
  First, we investigate the spatial correlator $G_\lambda(R)$ of the temporal link-variable $U_4$,
  and the relation to the instantaneous potential 
  $V_\lambda(R) \equiv -\frac{1}{a}{\rm ln} G_\lambda(R)$.
    For the large spatial separation of $R \equiv |\vec{x}-\vec{y}| \rightarrow \infty$,  
    $G_\lambda(R)$ behaves asymptotically as
    \begin{eqnarray}
      G_\lambda(R) &\equiv& 
      \langle \mathrm{Tr} \ [U_4^\dagger(\vec{x},t) U_4(\vec{y},t)] \rangle \nonumber \\
      &\rightarrow& 
        \langle (U_4)_{ij}^* \rangle
        \langle (U_4)_{ij} \rangle 
        = \frac{1}{3}
        \langle \mathrm{Tr} \ U_4 \rangle^2,
      \label{eqAsymUCorr}
    \end{eqnarray}
    where
    $\langle (U_4)_{ij}\rangle=\frac{1}{3}\langle \mathrm{Tr} \ U_4 \rangle \delta_{ij} 
    \in \mathbf{R}$ from the global color symmetry.
    Here, $\frac{1}{3}\langle \mathrm{Tr} \ U_4 \rangle^2$ is found to  
    give the lower bound of $G_\lambda(R)$.
    If $\langle \mathrm{Tr} \ U_4\rangle$ takes some finite value,
    $V_\lambda(R)$ inevitably saturates for large $R$. 
    Then, $\langle \mathrm{Tr} \ U_4\rangle=0$ is a necessary condition for 
    the deep-infrared confinement feature of 
    $V_\lambda(R=\infty) = \infty$.      

    In the Coulomb gauge,
    $\langle \mathrm{Tr} \ U_4 \rangle$ is zero due to the remnant symmetry, 
    as is shown in Appendix.
    Therefore, as $R \rightarrow \infty$, the correlator $G_\lambda(R)$ converges to zero, 
    and $V_\lambda(R)\equiv -\frac{1}{a}{\rm ln} G_\lambda(R) \rightarrow +\infty$, 
    which corresponds to the deep-infrared confinement.

    For the general case of $\lambda \neq 0$, however, 
    $\langle \mathrm{Tr} \ U_4 \rangle$ has a non-zero value, 
    and $G_\lambda(R)$ approaches to some finite constant.
    The finiteness of $\langle \mathrm{Tr} \ U_4 \rangle$
    gives a saturation of $V_\lambda(R)$,  
    which leads to the absence of its linear part
    in the case of rapid convergence. 

\begin{figure}
      \centering
      \includegraphics[width=8cm,clip]{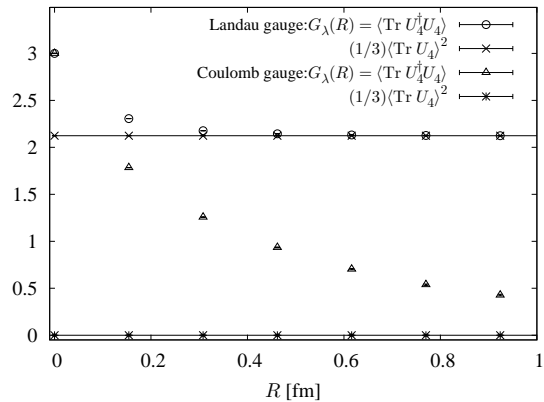}
      \caption{ \label{figU4Corr}
        The spatial correlator $G_\lambda(R) \equiv 
          \langle \mathrm{Tr} \ U_4^\dagger(\vec{x},a) U_4(\vec{y},a)\rangle$
        $(R = |\vec{x}-\vec{y}|)$  
        in the Landau gauge (open circles) and the Coulomb gauge (open triangles), 
        together with its asymptotic value of 
        $\frac{1}{3}\langle\mathrm{Tr} \ U_4 \rangle^2$ (solid lines and cross symbols).        
        In the Landau gauge, $\langle \mathrm{Tr} \ U_4\rangle \ne 0$, 
        and $G_\lambda(R)$ rapidly converges to a constant for $R \gtrsim$ 0.4fm,
        which leads to a rapid saturation of the instantaneous potential 
        $V_\lambda(R)\equiv -\frac{1}{a}{\rm ln} G_\lambda(R)$.
        In the Coulomb gauge, 
        $\langle \mathrm{Tr} \ U_4 \rangle = 0$, and
        $G_\lambda(R)$ decreases monotonically to zero as $R$, 
        which leads to $V_\lambda(R)=+\infty$ for $R=\infty$.
       }
    \end{figure}

    Figure \ref{figU4Corr} shows $G_\lambda(R)$ and its asymptotic value 
    $\frac{1}{3} \langle \mathrm{Tr} \ U_4\rangle^2$ 
    in the Landau and the Coulomb gauges.
    In the Landau gauge, $\langle \mathrm{Tr} \ U_4 \rangle$ has a large expectation value,
    according to the maximization of $\sum_x \mathrm{Re} \ \mathrm{Tr} \ U_\mu(x)$,
    and $G_\lambda(R)$ rapidly converges to a finite constant for $R \gtrsim$ 0.4fm, 
    which leads to a rapid saturation of the instantaneous potential $V_\lambda(R)$.
    In the Coulomb gauge, we find 
    $\langle \mathrm{Tr} \ U_4 \rangle = 0$,  
    and $G_\lambda(R)$ decreases monotonically to zero as $R$,
    which leads to $V_\lambda(R)=+\infty$ for $R=\infty$.

\begin{figure}
      \centering
      \includegraphics[width=8cm,clip]{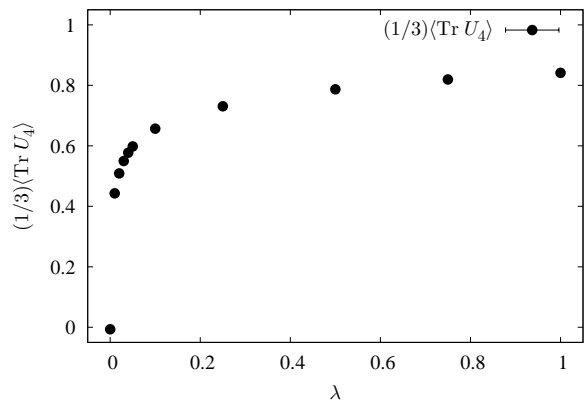}
      \caption{ \label{figTrU4}
        The expectation value of $\frac{1}{3}\langle \mathrm{Tr} \ U_4 \rangle$ in generalized Landau gauge.
        For $\lambda \ne 0$, 
        $\langle \mathrm{Tr} \ U_4 \rangle$ takes a non-zero value,
        and it approaches to zero continuously as $\lambda \rightarrow 0$.
        The value of $\langle \mathrm{Tr} \ U_4 \rangle$ relates to  
        the infrared behavior of the correlator $G_\lambda(R)$ 
        and the instantaneous potential $V_\lambda(R)$.
      }
    \end{figure}

    Figure \ref{figTrU4} shows $\lambda$-dependence of 
    $\frac{1}{3}\langle \mathrm{Tr} \ U_4 \rangle$ in generalized Landau gauge.
    For $\lambda \ne 0$, $\langle \mathrm{Tr} \ U_4 \rangle$ takes a non-zero real value, 
    and it approaches to zero continuously as $\lambda \rightarrow 0$.
    Here, it largely changes in the small region of $0 \le \lambda \lesssim 0.1$.
    The finiteness of $\langle \mathrm{Tr} \ U_4 \rangle$ is directly related to
    the infrared damping of the correlator $G_\lambda(R)$ 
    and the infrared form of the instantaneous potential $V_\lambda(R)$.

    \subsection{Asymptotic behavior of Polyakov-line correlator and $T$-length potential}
      Next, we consider 
      $T$-length terminated Polyakov-line correlator
      $G_\lambda(R,T)$,
      which behaves asymptotically as
      \begin{eqnarray} 
        G_\lambda(R,T) &\equiv& 
        \langle \mathrm{Tr} L^\dagger(\vec{x},T)L(\vec{y},T)\rangle \nonumber \\
        &\rightarrow& \frac{1}{3} \langle \mathrm{Tr} \ L(T) \rangle^2
      \end{eqnarray} 
      for large separation of $R = |\vec{x}-\vec{y}|$.
      As well as the instantaneous potential,
      $\frac{1}{3} \langle \mathrm{Tr} \ L(T)\rangle^2$ 
      is found to give the lower bound of the correlator $G_\lambda(R,T)$,
      and the finiteness of $\langle \mathrm{Tr} \ L(T) \rangle$ 
      is responsible for the infrared saturation of the finite-time potential $V_\lambda(R,T)$.

\begin{figure}
        \centering
        \includegraphics[width=8cm,clip]{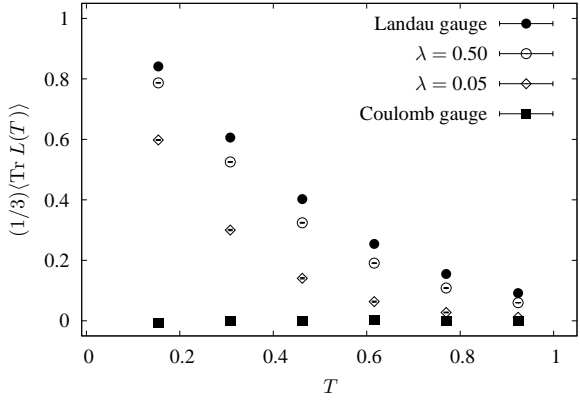}
        \caption{ \label{figTermPol}
          $T$-dependence of 
          $\frac{1}{3}\langle \mathrm{Tr} \ L(T) \rangle$ in generalized Landau gauge.
          In the Coulomb gauge, $\langle \mathrm{Tr} \ L(T) \rangle$ is always zero.
          For $\lambda \neq 0$,
          $\langle \mathrm{Tr} \ L(T) \rangle$ is a decreasing function of $T$,
          and it converges to zero in large-$T$ limit.
        }
     \end{figure}

      Figure \ref{figTermPol} shows $T$-dependence of
      the terminated Polyakov-line $\frac{1}{3}\langle \mathrm{Tr} \ L(T) \rangle$ 
      in generalized Landau gauge
      for typical values of $\lambda$.
      In the Coulomb gauge ($\lambda$=0), $\langle \mathrm{Tr} \ L(T) \rangle$ is always zero
      as well as $\langle \mathrm{Tr} \ U_4 \rangle$, 
      which means that $V_\lambda(R=\infty,T) = -\frac{1}{T}{\rm ln} G_\lambda(R=\infty,T)=+\infty$ 
      for any values of $T$.
      Actually, finite-time potential $V_\lambda(R,T)$ always has a linear part 
      in the Coulomb gauge, as shown in Fig.\ref{figExtendedPot}.

      For $\lambda \neq 0$,
      $\langle \mathrm{Tr} \ L(T) \rangle$ is a decreasing function of $T$,
      and it converges to zero in large-$T$ limit.
      At $T = N_t$,
      the $T$-length terminated Polyakov-line $\langle \mathrm{Tr} \ L(T) \rangle$ 
       results in the Polyakov loop, and $\langle \mathrm{Tr} \ L(N_t) \rangle = 0$ in the confinement phase.
      Therefore, $\langle \mathrm{Tr} \ L(T)\rangle$ converges to zero as $T \rightarrow N_t$, 
      and then one finds  
      \begin{eqnarray} 
      G_\lambda(R=\infty,N_t)&=&0, \\
      V_\lambda(R=\infty,N_t)&=& -\frac{1}{T}{\rm ln} G_\lambda(\infty,N_t)=+\infty,
      \end{eqnarray} 
      which gives a confinement potential.
      From Fig.\ref{figTermPol}, this convergence is found to be fast for smaller $\lambda$-value, 
      and such a convergence is closely related to the growing of finite-time string tension $\sigma_\lambda(T)$.

    \subsection{Gluon propagator 
    and instantaneous potential in the Landau gauge}
    In this subsection, we discuss the relation between the gluon propagator
    and the instantaneous potential in the Landau gauge.

    The gluon propagator is a two-point function
    of the gauge field $A_\mu(x)$, and is defined in Euclidean QCD as
    \begin{equation}
      D_{\mu\nu}(x,y) \equiv \langle \mathrm{Tr} A_\mu(x) A_\nu(y) \rangle,
    \end{equation}
    where the trivial color structure is dropped off by taking the trace.
    In the Landau gauge, we use the expression of $A_\mu$ in terms of $U_\mu$ as
    \begin{equation}
      A_4(x) = \frac{1}{2iag}[U_4(x)-U_4^\dagger(x)] + \mathcal{O}(a^2).
    \end{equation}
    Note that this expression is only justified 
    in the Landau gauge, or more generally in large-$\lambda$ gauges, 
    where the fluctuation of $A_4$ is highly suppressed.

    Then, the gluon propagator $D_{\mu\nu}$
    is expressed using link-variables, e.g.,  
    \begin{eqnarray}
      \label{eqPropInU}
      &&a^2g^2 D_{44}(x,y)  
      = a^2 g^2 \langle \mathrm{Tr} A_4(x)A_4(y)\rangle \nonumber \\
      &\simeq& - \frac{1}{4}\langle \mathrm{Tr} 
      [U_4(x)-U_4^\dagger(x)][U_4(y)-U_4^\dagger(y)]\rangle \nonumber \\
      &=& \langle \mathrm{Tr} [U_4(x)U_4^\dagger(y)]\rangle \nonumber \\
      && \quad   - \frac{1}{4} 
           \langle \mathrm{Tr}[U_4(x)+U_4^\dagger(x)]
                              [U_4(y)+U_4^\dagger(y)]\rangle. 
    \end{eqnarray}
    The last term in Eq.(\ref{eqPropInU}) has only
    $O(a^4)$-order $(x-y)$-dependence,
    and actually it changes only a few $\%$ in the Landau gauge at $\beta$=5.8, 
    so that we here approximate this term as a constant $C$. Thus, Eq.(\ref{eqPropInU}) reduces  
    \begin{equation}
      \label{eqPropInst}
      a^2 g^2 D_{44}(x,y) \simeq \langle \mathrm{Tr} [U_4(x) U_4^\dagger(y)] \rangle
      - C,
    \end{equation}
    and we calculate the instantaneous potential $V_{\rm inst}(R)$ as 
    \begin{eqnarray}
      \label{eqVinstD44}
      V_{\rm inst}(R) &=& -\frac{1}{a} 
      \ln 
      \langle \mathrm{Tr} [U_4(\vec{x},a) U_4^\dagger(\vec{y},a)] \rangle \nonumber \\
      &=& - \frac{1}{a} \ln \left[
      C + a^2 g^2 D_{44}(R) \right] \nonumber \\
      &\simeq& - \frac{a g^2}{C} D_{44}(R) + \mathrm{const.}
    \end{eqnarray}
    In this way, the instantaneous potential $V_{\rm inst}(R)$ is
    expressed by using the $44$-component of the gluon propagator
    in the Landau and large-$\lambda$ gauges.

    In the previous work \cite{Iritani09}, 
    we have found that the Landau-gauge gluon propagator is well reproduced 
    by the four-dimensional Yukawa-function as 
    \begin{equation}
      D(r) \equiv D_{\mu\mu}(r) \propto \frac{1}{r}e^{-mr},
    \end{equation}
    with the Yukawa mass-parameter $m \simeq$ 0.6GeV,
    in the region of $r = 0.1 \sim 1$ fm.
    Apart from a pre-factor from the tensor factor,
    $D_{44}(R)$ approximately behaves as the Yukawa-function,
    and therefore the instantaneous potential is expressed as
    \begin{equation}
      V_{\rm inst}(R) \simeq - \frac{A}{R}e^{-mR}
      + \mathrm{const.}
    \end{equation}
    in the Landau gauge.

    \begin{figure}
      \centering
      \includegraphics[width=8cm,clip]{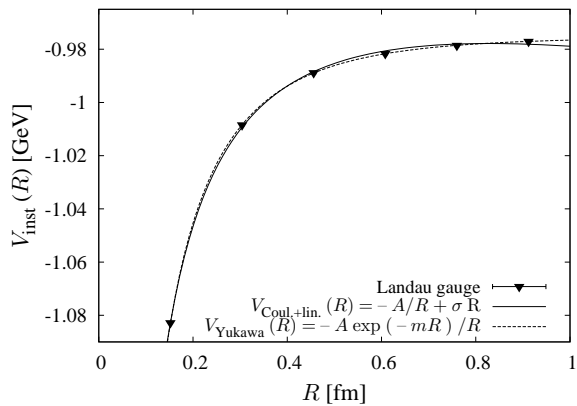}
      \caption{ \label{figInstLandau}
        Fit-result of the instantaneous potential $V_{\rm inst}(R)$ 
        in the Landau gauge
        using Coulomb plus linear form (solid line) 
        and Yukawa-function form (dashed line).
        Both forms well reproduce lattice QCD result.
        The best-fit parameter of the Yukawa mass is $m$ = 0.634(3)GeV,
        which coincides with the infrared effective gluon mass obtained from  
        the Landau-gauge gluon propagator \cite{Iritani09}.
      }
    \end{figure}

    Figure \ref{figInstLandau} shows the two fit-results of 
    instantaneous potential in the Landau gauge, 
    using the Coulomb plus linear form 
    $V_{\rm Coul.+lin.}(R) = -A/R + \sigma R$
    and the Yukawa-function form 
    $V_{\rm Yukawa}(R) = -A\exp(-mR)/R$.
    Both functions well reproduce the lattice QCD result.
    The best-fit Yukawa mass-parameter is $m$=0.634(3)GeV, 
    and this value coincides with the infrared effective gluon mass 
    obtained from the Landau-gauge gluon propagator \cite{Iritani09}.

    We note again that this relation is only valid in the Landau and
    large-$\lambda$ gauges, where
    the temporal link-variable $U_4$ can be expanded in terms of lattice spacing $a$,
    and the last-term in Eq.(\ref{eqPropInU}) is almost constant.

\section{Summary and Discussion}
  In this paper, aiming to grasp the gauge dependence of gluon properties,
  we have investigated generalized Landau gauge and applied it to 
  instantaneous interquark potential in SU(3) quenched lattice QCD at $\beta$=5.8.
  In the Coulomb gauge, the instantaneous potential is expressed by 
  the sum of Coulomb potential and linear potential 
  with 2-3 times larger string tension. 
  In contrast, the instantaneous potential has no linear part 
  in the Landau gauge. Thus, there is a large gap between these two gauges.
  Using generalized Landau gauge, we have found that the instantaneous 
  potential $V_\lambda(R)$ is connected continuously from the Landau gauge 
  towards the Coulomb gauge, and the linear part in $V_\lambda(R)$ 
  grows rapidly in the neighborhood of the Coulomb gauge.
  
  Since the slope $\sigma_\lambda$ of the instantaneous potential 
  $V_\lambda(R)$ grows continuously from 0 to 2-3$\sigma_{\rm phys}$, 
  there must exist some specific intermediate gauge 
  where the slope $\sigma_\lambda$ coincides with 
  the physical string tension $\sigma_{\rm phys}$. 
  From the lattice QCD calculation, the specific $\lambda$-parameter,
  $\lambda_C$, is estimated to be about $0.02$.
  In this $\lambda_C$-gauge, the physical static interquark potential 
  $V_{\rm phys}(R)$ is approximately reproduced 
  by the instantaneous potential $V_\lambda(R)$. 

  We have also investigated $T$-length terminated Polyakov-line correlator, 
  and its corresponding finite-time potential $V_\lambda(R,T)$, 
  which is a generalization of the instantaneous potential $V_\lambda(R)$, 
  in generalized Landau gauge.
  The behavior of the slope $\sigma_\lambda(T)$ of the finite-time potential 
  is classified into three groups: the Coulomb-like gauge 
  ($0 \le \lambda \lesssim 0.01$), the Landau-like gauge ($0.1 \lesssim \lambda \le 1$), 
  and neighborhood of $\lambda_C$-gauge ($\lambda \sim \lambda_C \simeq 0.02$).
  In the Coulomb-like gauge, the slope $\sigma_\lambda(T)$ 
  is a decreasing function of $T$, and seems to approach to 
  physical string tension $\sigma_{\rm phys}$ for large $T$.
  In the Landau-like gauge, $\sigma_\lambda(T)$ is an increasing function.
  Around the $\lambda_C$-gauge, 
  $\sigma_\lambda(T)$ has a weak $T$-length dependence.
  We have also investigated $T$-length terminated Polyakov-line correlator 
  and its relation to the finite-time potential.

  Finally, we consider a possible gauge of QCD to describe 
  the quark potential model from the viewpoint of instantaneous potential. 
  The quark potential model is a successful nonrelativistic framework 
  with a potential instantaneously acting among quarks, 
  and describes many hadron properties in terms of quark degrees of freedom. 
  In this model, there are no dynamical gluons, and gluonic effects 
  indirectly appear as the instantaneous interquark potential.

  As for the Coulomb gauge, the instantaneous potential has too large 
  linear part, which gives an upper bound on the static potential 
  \cite{Zwanziger03}. It has been suggested by Greensite et al. that
  the energy of the overconfining state is lowered by inserting 
  dynamical gluons between (anti-)quarks, 
  which is called ``gluon-chain picture''. 
  This gluon-chain state is considered as the ground-state 
  in the Coulomb gauge \cite{Greensite03,Greensite04,GreensiteThorn}.
  Therefore, dynamical gluon degrees of freedom must be 
  also important to describe hadron states in the Coulomb gauge. 

\begin{figure}
  \centering
  \includegraphics[width=8cm,clip]{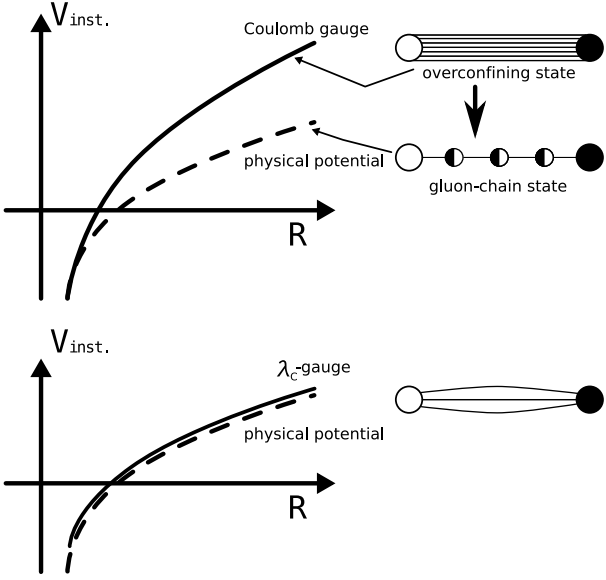}
  \caption{\label{figGluonChain}
  The schematic illustration of the Coulomb gauge and
  the $\lambda_C$-gauge.
  In the Coulomb gauge, the instantaneous potential
  gives an ``overconfining state'' as an excited-state 
  for the quark-antiquark system.
  The ground-state is considered as the gluon-chain state,
  which contains dynamical gluons between static sources.
  In the $\lambda_C$-gauge, the instantaneous potential
  gives the physical interquark potential approximately,
  and dynamical gluons need not appear.
  }
\end{figure}

  For $\lambda_C$-gauge, the physical interquark potential
  $V_{\rm phys}(R)$ is approximately reproduced 
  by the instantaneous potential $V_{\lambda_C}(R)$ unlike the Coulomb gauge, 
  as schematically shown in Fig.\ref{figGluonChain}.
  This physically means that all other complicated effects 
  including dynamical gluons and ghosts are approximately cancelled
  in the $\lambda_C$-gauge, and therefore we do not need to 
  introduce any redundant gluonic degrees of freedom. 
  The absence of dynamical gluon degrees of freedom 
  would be a desired property for the quark model picture. 
  The weak $T$-length dependence of $\sigma_\lambda(T)$ 
  around the $\lambda_C$-gauge ($T$-length stability) 
  is also a suitable feature for the potential model.

  In this way, as an interesting possibility, 
  the $\lambda_C$-gauge is expected to be a useful gauge 
  in considering the linkage from QCD to the quark potential model.
  In addition, since $\lambda_C \simeq 0.02 \ll 1$ 
  is a very small parameter in this framework, 
  it is interesting to apply  
  the perturbative technique in terms of $\lambda_C$ 
  for the calculation of the Faddeev-Popov determinant and so on.

\begin{acknowledgments}
  The authors thank Dr. H. Iida for his useful arguments.
  H.S. is supported in part by the Grant for Scientific
  Research [(C) No.19540287, Priority Areas ``New Hadrons'' (E01:21105006)] 
  from the Ministry of Education,
  Culture, Science and Technology (MEXT) of Japan.
  This work is supported by the Global COE Program,
  ``The Next Generation of Physics, Spun from Universality and Emergence".
  The lattice QCD calculations have been done on NEC-SX8 at
  Osaka University.
\end{acknowledgments}

\appendix
\section{On the Coulomb gauge}
  In this appendix, we briefly discuss the difference between the Coulomb gauge
  and the $\lambda \rightarrow 0$ limit of generalized Landau gauge.
  
  \subsection{Residual gauge degrees of freedom in the Coulomb gauge}
    We comment on the residual gauge degrees of
    freedom in the Coulomb gauge.
    In lattice QCD,
    the Coulomb gauge fixing condition is expressed by
    the maximization of the quantity
    \begin{equation}
      R_{\rm Coul}[U] \equiv \sum_{\vec{x},t} \sum_{i=1}^3
      \mathrm{Re} \ \mathrm{Tr} \ U_i(\vec{x},t)
    \end{equation}
    by the gauge transformation.

    Now, we consider the spatially-global gauge transformation as
    \begin{eqnarray}
      U_i(\vec{x},t) &\rightarrow & \Omega(t) U_i(\vec{x},t) \Omega^\dagger(t), \\
      U_4(\vec{x},t) &\rightarrow & \Omega(t) U_4(\vec{x},t) \Omega^\dagger(t+1),
    \end{eqnarray}
    with the gauge function $\Omega(t) \in {\rm SU}(N_c)$.
    $R_{\rm Coul}[U]$ is invariant under this transformation,
    \begin{eqnarray}
      R_{\rm Coul}[U] &\equiv& 
        \sum_{\vec{x},t} \sum_{i=1}^3 \mathrm{Re} \ \mathrm{Tr} \ U_i(\vec{x},t) \nonumber \\
      &\rightarrow& 
        \sum_{\vec{x},t} \sum_{i=1}^3 \mathrm{Re} \ \mathrm{Tr} \
          \Omega(t) U_i(\vec{x},t) \Omega^\dagger(t) \nonumber \\
      &=& \sum_{\vec{x},t} \sum_{i=1}^3 \mathrm{Re} \ \mathrm{Tr} \ U_i(\vec{x},t).
    \end{eqnarray}
    Therefore, the Coulomb gauge has the corresponding residual gauge degrees of freedom.

    Under this residual symmetry, however,  
   $\mathrm{Tr} \ U_4$ is gauge-variant as 
    \begin{eqnarray}
      \mathrm{Tr} \ U_4(\vec{x},t) &\rightarrow&
        \mathrm{Tr} \ \Omega(t) U_4(\vec{x},t) \Omega^\dagger(t+1) 
    \end{eqnarray}
    so that the expectation value of $\mathrm{Tr} \ U_4$ is to be zero
    in the Coulomb gauge.

    In the generalized Landau gauge with non-zero $\lambda$-parameter,
    this residual symmetry does not exist,
    and hence $\mathrm{Tr} \ U_4$ has a finite expectation value. 
    (See Fig.\ref{figTrU4}.)

  \subsection{Convergence into Coulomb gauge}
    We here investigate the convergence of the 
    generalized Landau gauge into the Coulomb gauge in the limit of $\lambda \rightarrow 0$.
    To check the convergence, we evaluate the quantity $\langle (\partial_i A_i^a)^2 \rangle$,
  \begin{figure}[h]
    \centering
    \includegraphics[width=8cm,clip]{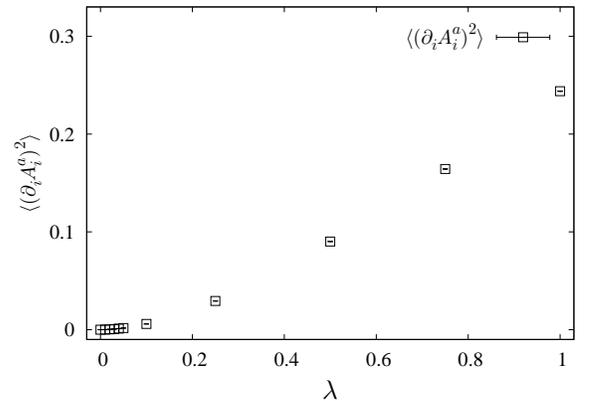}
    \caption{\label{figConvergence}
      The lattice QCD result of $\langle (\partial_i A_i^a)^2 \rangle$ in the generalized Landau gauge.
      As $\lambda \rightarrow 0$, this quantity
      goes to zero monotonically.
    }
  \end{figure}
    \begin{eqnarray} 
      \langle \left(\partial_i A_i^a\right)^2 \rangle   
      &\equiv &
      \frac{1}{(N_c^2-1){N_{\rm site}}} 
\nonumber \\
      &\times &
      \sum_{x=1}^{N_{\rm site}} \sum_{a=1}^{N_c^2-1} 
      \Big\{ \sum_{i=1}^3 \left[ A_i^a(x) - A_i^a(x-\hat{i})\right]\Big\}^2.~~~~
    \end{eqnarray}
    In the Coulomb gauge, $\langle (\partial_i A_i^a)^2 \rangle$ is equal to zero.

    Figure \ref{figConvergence} shows the lattice QCD result of $\langle (\partial_i A_i^a)^2 \rangle$,
    which is monotonically decreasing toward zero by varying $\lambda$ from 1 to 0. 
    This result supports that the generalized Landau gauge approaches 
    to the Coulomb gauge in the $\lambda \rightarrow 0$ limit.

\end{document}